\documentstyle[amstex,amssymb,preprint,aps]{revtex}

\def\y{\eta}

\newcommand{\be}{\begin{equation}}
\newcommand{\ee}{\end{equation}}
\newcommand{\bn}{\begin{eqnarray}}
\newcommand{\en}{\end{eqnarray}}

\newcommand{\ba}{\begin{array}}
\newcommand{\ea}{\end{array}}

\begin{document}
\draft
\title{{\Large Quantization of spinning particle in arbitrary background}}
\author{S.P. Gavrilov\thanks{%
Universidade Federal de Sergipe, Brazil; on leave from Tomsk Pedagogical
University, Russia; present e-mail: gavrilov\symbol{64}ufs.br} and D.M.
Gitman\thanks{%
e-mail: gitman\symbol{64}fma.if.usp.br}}
\address{Instituto de F\'{\i}sica, Universidade de S\~ao Paulo,\\
C.P. 66318, 05315-970 S\~ao Paulo, SP, Brasil}
\date{\today}
\maketitle

\begin{abstract}
The article is a natural continuation of the papers by Gavrilov and Gitman
(Class.Quant.Grav. {\bf 17} (2000) L133; Int. J. Mod. Phys. A15 (2000) 4499)
devoted to relativistic particle quantization. Here we generalize the
problem, considering the quantization of a spinning particle in arbitrary
gravitational background. The nontriviality of such a generalization is
related to the neccessity of solving complicated ordering problems. Similar
to the flat space-time case, we show in the course of the canonical
quantization how a consistent relativistic quantum mechanics of spinning
particle in gravitational and electromagnetic backgrounds can be constructed.
\end{abstract}

\section{Introduction}

The problem of quantizing the classical (pseudoclassical) models of
relativistic particles was discussed in numerous articles \cite{1,2}. In our
last publications \cite{GavGi00}, we presented a new solution of this
problem, which shows how a consistent relativistic quantum mechanics can be
obtained in the course of the canonical quantization of relativistic
particle models. We stress that the new construction gives a solution to the
old problem of how to construct a consistent quantum mechanics on the base
of a relativistic wave equation. The quantization was performed for a
spinless particle in arbitrary electromagnetic and gravitational
backgrounds, as well as for a spinning particle, but only in an external
electromagnetic background. The case of the spinning particle in curved
space-time was not considered in the above publications. In the present
article we discuss in detail the canonical quantization of spinning particle
moving in arbitrary electromagnetic and gravitational backgrounds in $3+1$
dimensions. Here we meet both technical and conceptual problems, in
particular the ordering problem. It is enough to mention that quantization
in the even simpler corresponding nonrelativistic particle case is an open
problem. It attracts attention up to the present and several points of view
have been brought up on its solution \cite{DeWit52}. The relativistic case,
which naturally absorbs all known difficulties of its nonrelativistic
analog, is essentially richer and more complicated due to its gauge nature.

\section{Pseudoclassical model of a spinning particle in curved space-time}

A pseudoclassical action of a spin one-half relativistic particle in $3+1$
dimensions, with spinning degrees of freedom describing by anticommuting
(Grassmann) variables, was discussed in \cite{1,2}. In flat space-time ($%
g_{ab}=\eta _{ab}={\rm diag}(1,-1,-1,-1),~a,b={0,1,2,3,\;}$Latin letters
from the beginning of the alphabet are used for the Lorentz indices$),$ and
in the presence of an external electromagnetic field $A_{a}\left( x\right) ,$
the action\ can be written in the following Lorentz invariant form: 
\begin{align}
& S=\int_{0}^{1}Ld\tau \,,\;L=-\frac{\eta _{ab}}{2e}\left( \dot{x}^{a}-i\xi
^{a}\chi \right) \left( \dot{x}^{b}-i\xi ^{b}\chi \right) -\frac{e}{2}m^{2}-q%
\dot{x}^{a}A_{a}\left( x\right)  \nonumber \\
& \;+iqeF_{ab}\left( x\right) \xi ^{a}\xi ^{b}-i\xi _{a}\dot{\xi}^{a}+i\xi
_{4}\dot{\xi}_{4}+im\xi _{4}\chi \,,  \label{1}
\end{align}
where the$\;$coordinates $x^{a}$ of the particle and the variable$\ e$ are
Grassmann-even; the Lorentz vector $\xi ^{a}$, the (pseudo) scalar $\xi
_{4}\,,\,\ $and the scalar$\ \chi $ are Grassmann-odd. All the variables
depend on the parameter $\tau \in \lbrack 0,1]$, which plays here the role
of time. Dots above the variables denote their derivatives with respect to $%
\tau $. There are two types of gauge transformations under which the action (%
\ref{1}) is invariant: the reparametrizations and the supertransformations 
\begin{eqnarray}
&&\delta x^{a}={\dot{x}}^{a}\varepsilon \,,\;\delta e=d\left( e\varepsilon
\right) /d\tau ,\;\delta \xi ^{a}=\dot{\xi}^{a}\varepsilon \;,\;\delta \xi
_{4}=\dot{\xi}_{4}\varepsilon \,,\;\delta {\chi }=d\left( \chi \varepsilon
\right) /d\tau \,;  \nonumber \\
&&\delta x^{a}=i\xi ^{a}\epsilon \,,~\delta e=i\chi \epsilon \,,\;\delta
\chi =\dot{\epsilon}\,,\,\delta \xi ^{a}=\left( \dot{x}^{a}-i\xi ^{a}\chi
\right) \epsilon /2e\,,~\delta \xi _{4}=-m\epsilon /2\,,  \label{1a}
\end{eqnarray}
where $\varepsilon (\tau )\,$and$\ \epsilon (\tau )$ are $\tau $-dependent
gauge parameters, the first one is even and the second one is odd.

We generalize the action (\ref{1}) to the curved space-time case (without
torsion\footnote{%
Introduction of the interaction with a torsion field in the model was
discussed in \cite{GeyGiS00}}) with a metric tensor $g_{\mu \nu }\left(
x\right) $ as follows\footnote{%
A different (nonsupersymmetric) form of the spinning particle action in
curved space time was considered in \cite{3}. That action follows from (\ref
{2}) in a special gauge.}: 
\begin{align}
S& =\int_{0}^{1}Ld\tau \,,\ L=-\frac{g_{\mu \nu }\left( x\right) }{2e}\left[ 
\dot{x}^{\mu }-i\xi ^{\mu }\left( x\right) \chi \right] \left[ \dot{x}^{\nu
}-i\xi ^{\nu }\left( x\right) \chi \right] -\frac{e}{2}m^{2}-q\dot{x}^{\mu
}A_{\mu }\left( x\right)  \nonumber \\
& +iqeF_{\mu \nu }\left( x\right) \xi ^{\mu }\left( x\right) \xi ^{\nu
}\left( x\right) -i\xi _{\mu }\left( x\right) D_{\tau }\xi ^{\mu }\left(
x\right) +i\xi _{4}\dot{\xi}_{4}\;+im\xi _{4}\chi .  \label{2}
\end{align}
Here $\xi ^{\mu }(x)=e_{a}^{\mu }(x)\xi ^{a}$ are world vectors (Greek
letters are used for the world indices, e.g. $\mu =0,1,2,3),$ and $\xi
^{a}\, $are Lorentz vectors$,$ where $e_{a}^{\mu }(x)$ is the vierbein field 
\cite{Kibbl61}, and $D_{\tau }$ is the covariant derivative with respect to $%
\tau ,$%
\begin{align}
& D_{\tau }\xi ^{\mu }\left( x\right) =\xi _{;\sigma }^{\mu }\left( x\right)
\,\dot{x}^{\sigma }+e_{a}^{\mu }(x)\dot{\xi}^{a},\;D_{\tau }\xi ^{a}=\dot{\xi%
}^{a}+\xi _{b}\omega _{\nu }^{ba}\left( x\right) \dot{x}^{\nu }=e_{\mu
}^{a}(x)D_{\tau }\xi ^{\mu }\left( x\right) ,  \nonumber \\
\;& \,e_{\mu }^{a}(x)e_{a\nu }(x)=g_{\mu \nu }\left( x\right) ,\;e_{a}^{\mu
}(x)e_{b\mu }(x)=\eta _{ab},\ \xi _{;\sigma }^{\mu }\left( x\right)
=\partial _{\sigma }\xi ^{\mu }\left( x\right) +\Gamma _{\nu \sigma }^{\mu
}\left( x\right) \xi ^{\nu }\left( x\right) .  \label{3}
\end{align}
Here$\ \omega _{\nu }^{ab}(x)=\left[ \partial _{\nu }e^{a\lambda
}(x)+e^{a\sigma }(x)\Gamma _{\sigma \nu }^{\lambda }(x)\right] e_{\lambda
}^{b}(x)$ are spin connections for the torsion-free case,$\ \omega _{\nu
}^{ab}\ =-\omega _{\nu }^{ba}$, and $\Gamma _{\nu \sigma }^{\mu }$ is the
affine connection. This action (\ref{2}) is invariant under general
coordinate transformations and is invariant under the gauge transformations 
\begin{eqnarray}
&&\delta x^{\mu }={\dot{x}}^{\mu }\varepsilon \;,\;\delta e=d\left(
e\varepsilon \right) /d\tau \;,\;\delta \xi ^{a}=\dot{\xi}^{a}\varepsilon
\;,\;\delta \xi _{4}=\dot{\xi}_{4}\varepsilon \;,\;\delta {\chi }=d\left(
\chi \varepsilon \right) /d\tau \,;  \nonumber \\
&&\delta x^{\mu }=i\xi ^{\mu }\epsilon ,~\delta e=i\chi \epsilon ,\;\delta
\chi =\dot{\epsilon}\,,\,\delta \xi ^{\mu }=\left( \dot{x}^{\mu }-i\xi ^{\mu
}\chi \right) \epsilon /2e,~\delta \xi _{4}=-m\epsilon /2,  \label{3a}
\end{eqnarray}

For the purpose of quantization, we select a reference frame which admits a
time synchronization over all space. Such a reference frame corresponds to a
special gauge $g_{0i}=0$ for which $g^{00}=g_{00}^{-1},\;g^{ik}g_{kj}=\delta
_{j}^{i}$. Such a reference frame always exists for any real space-time.
Besides, we choose a special gauge for the vierbein, $e_{0}^{a}(x)=\delta
_{0}^{a}\sqrt{g_{00}(x)},\;e_{i}^{0}(x)=0.$ Then, $\omega _{0}^{\bar{a}0}=0,$
$\xi _{\bar{a}}\xi ^{k}\omega _{k}^{\bar{a}0}=0$. In our index conventions,
barred Latin letters from the beginning of the alphabet denote spatial
Minkowski vectors, $a=(0,\bar{a}),$ $\bar{a}={1,2,3};~$Latin letters from
the middle of the alphabet represent spatial world vectors, so that $\mu
=(0,i),$ $i={1,2,3.}$

\section{Hamiltonian structure of the theory}

The expressions for the canonical momenta have the form 
\begin{eqnarray}
p_{\mu } &=&\frac{\partial L}{\partial \dot{x}^{\mu }}=-e^{-1}g_{\mu \nu
}\left( \dot{x}^{\nu }-i\xi ^{\nu }\chi \right) -qA_{\mu }\,-i\xi _{a}\xi
_{b}\omega _{\mu }^{ba},\;P_{e}=\frac{\partial L}{\partial \dot{e}}=0\,,\; 
\nonumber \\
\pi _{a} &=&\frac{\partial _{r}L}{\partial \dot{\xi}^{a}}=-i\xi _{a}\,,\;\pi
_{4}=-\frac{\partial _{r}L}{\partial \dot{\xi}_{4}}=-i\xi _{4}\,,\;P_{\chi }=%
\frac{\partial _{r}L}{\partial \dot{\chi}}=0\,.  \label{4}
\end{eqnarray}
They imply the following primary constraints ${\phi }_{B}^{(1)}=0,%
\;B=1,2,(3,n),\;n=0,1,2,3,4\,.$\ 
\begin{equation}
\phi _{1}^{(1)}=P_{\chi }\,,\;\phi _{2}^{(1)}=P_{e}\,,\;\phi
_{3,n}^{(1)}=\pi _{n}+i\xi _{n}\,.  \label{5}
\end{equation}
The total Hamiltonian \cite{Dirac64} has the form $H^{(1)}=H+\lambda
^{B}\phi _{B}^{(1)},$ where 
\begin{eqnarray}
H &=&-\frac{e}{2}\left[ (p_{\mu }+qA_{\mu }+i\xi _{a}\xi _{b}\omega _{\mu
}^{ba})g^{\mu \nu }(p_{\nu }+qA_{\nu }+i\xi _{c}\xi _{d}\omega _{\nu
}^{dc})-m^{2}+2iqF_{\mu \nu }\xi ^{\mu }\xi ^{\nu }\right]  \nonumber \\
&&+i\left[ (p_{\mu }+qA_{\mu }+i\xi _{a}\xi _{b}\omega _{\mu }^{ba})\xi
^{\mu }-m\xi _{4}\right] \chi \;.  \label{6}
\end{eqnarray}
Using the consistency conditions $\dot{\phi}^{(1)}=0$ for the primary
constraints, we find the secondary constraints $\phi ^{(2)}=0$, 
\begin{eqnarray}
\phi _{1}^{(2)} &=&(p_{\mu }+qA_{\mu }+i\xi _{a}\xi _{b}\omega _{\mu
}^{ba})\xi ^{\mu }+m\xi ^{4}\,,\;\;  \nonumber \\
\phi _{2}^{(2)} &=&(p_{\mu }+qA_{\mu }+i\xi _{a}\xi _{b}\omega _{\mu
}^{ba})g^{\mu \nu }(p_{\nu }+qA_{\nu }+i\xi _{c}\xi _{d}\omega _{\nu
}^{dc})-m^{2}+2iqF_{\mu \nu }\xi ^{\mu }\xi ^{\nu }\;,  \label{7}
\end{eqnarray}
and determine the $\lambda $'s, which correspond to the primary constraints $%
\phi _{3,n}^{(1)}$. No more secondary constraints arise from the consistency
conditions, and the $\lambda $'s that correspond to the constraints $\phi
_{1}^{(1)},\;\phi _{2}^{(1)}$ remain undetermined. The Hamiltonian $H$ is
proportional to the constraints, $H=-\frac{e}{2}\phi _{2}^{(2)}+i\phi
_{1}^{(2)}\chi \;.$

It is convenient to replace the initial set of constraints $\phi ^{(1)},\phi
^{(2)}$ by an equivalent one, which we define below. To this end we define
the principal value of the square root of an expression containing Grassmann
variables as the one which is positive whenever all generating elements of
the Grassmann algebra are set to zero. Suppose $r,$ 
\begin{equation}
\;r=\sqrt{g_{00}[m^{2}-\left( p_{k}+{\cal A}_{k}\right) g^{kl}\left( p_{l}+%
{\cal A}_{l}\right) +2qF_{\mu \nu }\xi ^{\mu }\pi ^{\nu }]},\;{\cal A}_{\mu
}=qA_{\mu }-\pi _{a}\xi _{b}\omega _{\mu }^{ba},  \label{7a}
\end{equation}
is such a principal value of the expression indicated. Then we introduce a
set of constraints $\phi ^{(1)},T$, equivalent to $\phi ^{(1)},\phi ^{(2)},$
where $\ $ 
\begin{eqnarray}
\ &&T_{1}=\left( p_{\mu }+qA_{\mu }+i\xi _{a}\xi _{b}\omega _{\mu
}^{ba}\right) \left( \pi ^{\mu }-i\xi ^{\mu }\right) -m\left( \pi _{4}-i\xi
_{4}\right) +2i\xi _{a}\left( \pi _{b}+i\xi _{b}\right) \omega _{\mu
}^{ba}\xi ^{\mu }\;,  \nonumber \\
\ &&T_{2}=p_{0}+{\cal A}_{0}+\zeta r,\;\,\zeta =\pm 1\,.  \label{8}
\end{eqnarray}
To check that, it is useful to take into account the following relation$\;$ 
\[
\phi _{2}^{(2)}=(-2\zeta r+T_{2})g^{00}T_{2}-\frac{i}{2}\phi
_{3,a}^{(1)}\left\{ \phi _{3,b}^{(1)}\eta ^{ba},\phi _{2}^{(2)}\right\} -%
\frac{i}{2}\phi _{3,a}^{(1)}2i\xi _{b}\omega _{\mu }^{ba}g^{\mu \nu }\frac{i%
}{2}\phi _{3,c}^{(1)}2i\xi _{d}\omega _{\nu }^{dc}. 
\]
In fact the constraint $T_{2}=0$ is a linearized analog of the quadratic
primary constraint $\phi _{2}^{(2)}=0$ (compare with the flat space-time
case \cite{GavGi00}). We can regard the discrete variable $\zeta =\pm 1$ as
the sign of the Grassmann valued quantity $p_{0}+{\cal A}_{0}\,,$%
\begin{equation}
\zeta =-{\rm sign}\left[ p_{0}+{\cal A}_{0}\right] \,,  \label{10}
\end{equation}
is an analog of the charge sign variable, well known in the flat-space case 
\cite{GavGi00} and especially important for all further constructions. One
can easily see from the equations (\ref{4}) that, similar to scalar particle
case,$\;{\rm sign}(\dot{x}^{0})=\zeta .$

The new set of constraints $\phi ^{(1)},T$ is explicitly divided in a subset
of the first-class constraints $\phi _{1}^{(1)},\phi _{2}^{(1)},T$, and in a
subset of second-class constraints $\phi _{3,n}^{(1)}$ . Indeed, 
\begin{equation}
\left\{ \phi _{\varkappa }^{(1)},\phi ^{(1)}\right\} =\left\{ \phi
_{\varkappa }^{(1)},T\right\} =\left. \left\{ T,\phi _{3,n}^{(1)}\right\}
\right| _{\phi =T=0}=\left. \left\{ T,T\right\} \right| _{\phi
=T=0}=0\;,\;\;\varkappa =1,2\,.  \label{9}
\end{equation}

Similar to the flat space-time case we impose first two gauge conditions $%
\phi ^{G}=0$, 
\begin{equation}
\phi _{1}^{G}=\pi ^{0}-i\xi ^{0}-\zeta \left( \pi _{4}-i\xi _{4}\right)
,\;\;\phi _{2}^{G}=x^{0}-\zeta \tau \;\;.  \label{11}
\end{equation}
>From the consistency conditions $\dot{\phi}_{1,2}^{G}=0,$ we find two
additional constraints 
\begin{equation}
\phi _{3}^{G}=\chi +2\zeta \alpha \Delta ^{-1}=0\;,\;\phi _{4}^{G}=e-g_{00}%
\tilde{\omega}^{-1}\left[ 1-\alpha \Delta ^{-1}\left( \pi ^{a}-i\xi
^{a}\right) e_{a}^{0}\right] =0,  \label{12}
\end{equation}
where

\begin{eqnarray}
&&\alpha =\frac{i\zeta g_{00}}{r+\tilde{\omega}}\left[ 2\left( \pi _{\bar{b}%
}-i\xi _{\bar{b}}\right) \omega _{k}^{\bar{b}0}g^{kl}\left( p_{l}+\widetilde{%
{\cal A}}_{l}\right) -2qF_{k\mu }e_{0}^{\mu }\left( \pi ^{k}-i\xi
^{k}\right) \right] \,,  \nonumber \\
&&\Delta =2\left[ \zeta (\tilde{\omega}_{0}+m)\right] \,,\;\widetilde{{\cal A%
}}_{\mu }=qA_{\mu }-\pi _{\bar{a}}\xi _{\bar{b}}\omega _{\mu }^{\bar{b}\bar{a%
}},\;\tilde{\omega}=\left. r\right| _{\pi _{0}=\bar{\pi}_{0},\;\xi _{0}=i%
\bar{\pi}_{0}}=\sqrt{g_{00}\left( \tilde{\omega}_{0}^{2}+\tilde{\rho}\right) 
}\;,  \nonumber \\
&&\tilde{\omega}_{0}=\sqrt{\left[ m^{2}-\left( p_{k}+\widetilde{{\cal A}}%
_{k}\right) g^{kl}\left( p_{l}+\widetilde{{\cal A}}_{l}\right) +2qF_{kl}\xi
^{k}\pi ^{l}\right] }\,,\;  \nonumber \\
&&\tilde{\rho}=2\left[ qF_{k\mu }e_{0}^{\mu }\left( \xi ^{k}+i\pi
^{k}\right) \bar{\pi}^{0}+\bar{\pi}_{0}\left( \xi _{\bar{b}}+i\pi _{\bar{b}%
}\right) \omega _{k}^{\bar{b}0}g^{kl}\left( p_{l}+\widetilde{{\cal A}}%
_{l}\right) \right] ,  \nonumber \\
&&\bar{\pi}_{0}=\Delta ^{-1}\left[ \left( p_{l}+\widetilde{{\cal A}}%
_{l}\right) \left( \pi ^{l}-i\xi ^{l}\right) +2i\xi _{\bar{a}}\left( \pi _{%
\bar{b}}+i\xi _{\bar{b}}\right) \omega _{j}^{\bar{b}\bar{a}}\xi ^{j}\right]
\,.  \label{13}
\end{eqnarray}
Then, from the consistency conditions $\dot{\phi}_{3,4}^{G}=0$ we can find $%
\lambda ^{1,2}$ . All the constraints $\left( \phi ^{(1)},T,\phi ^{G}\right) 
$ are already of second-class. As in flat-space case we pass from these
constraints to an equivalent set of second-class constraints $\Phi
_{a},\;a=1,2,...,13,$ 
\begin{eqnarray}
\Phi _{1} &=&t_{1}T_{1}+t_{2}T_{2}+f_{1}\phi _{1}^{G}+f_{0}\phi
_{3,0}^{(1)}+f_{\bar{a}0}\phi _{3,\bar{a}}^{(1)}\phi _{3,0}^{(1)}=p_{0}+%
\widetilde{{\cal A}}_{0}+\zeta \tilde{\omega}\,,\;\Phi _{2}=\phi
_{2}^{G},\;\Phi _{3}=\phi _{3,1}^{(1)}\,,\;  \nonumber \\
\Phi _{4} &=&\phi _{3,2}^{(1)}\,,\;\,\Phi _{5}=\phi _{3,3\;}^{(1)},\;\Phi
_{6}=T_{1}+bT_{2}+c\phi _{2}^{G},\;\Phi _{7}=\phi _{1}^{G}\,,\;\Phi
_{8}=\phi _{3}^{G}+d\phi _{2}^{G}+v\phi _{1}^{G}+u\;\phi _{2}^{\left(
1\right) },\;  \nonumber \\
\ \Phi _{9} &=&\phi _{1}^{(1)}\,,\;\Phi _{10}=\phi _{4}^{G}+w\phi
_{2}^{G}+z\Phi _{7}+s\Phi _{6}\,,\;\Phi _{11}=\phi _{2}^{(1)}\,,\;\Phi
_{12}=\phi _{3,0}^{\left( 1\right) },\ \Phi _{13}=\phi _{3,4}^{(1)}\,\,,
\label{14}
\end{eqnarray}
where 
\begin{eqnarray*}
&&t_{1}=-\alpha \Delta ^{-1},\;t_{2}=1+\alpha \Delta ^{-1}\left( \pi
^{a}-i\xi ^{a}\right) e_{a}^{0},\;f_{1}=\zeta m\alpha \Delta
^{-1},\;f_{0}=\alpha \zeta \left( rg_{00}^{-1/2}+m\right) , \\
\; &&f_{\bar{a}0}=\alpha \Delta ^{-1}\xi _{\bar{b}}\omega _{0}^{\bar{b}\bar{a%
}}g_{00}^{-1/2},\;b=-\frac{\left\{ \phi _{2}^{G},T_{1}\right\} }{\left\{
\phi _{2}^{G},T_{2}\right\} }\,,\;c=-\frac{\left\{ T_{1}+bT_{2},\Phi
_{1}\right\} }{\left\{ \phi _{2}^{G},\Phi _{1}\right\} }\,,\;u=-\frac{%
\left\{ \phi _{3}^{G}+v\phi _{1}^{G},\phi _{4}^{G}\right\} }{\left\{ \phi
_{2}^{(1)},\phi _{4}^{G}\right\} }\,, \\
&&v=-\frac{\left\{ \phi _{3}^{G},\Phi _{6}\right\} }{\left\{ \Phi _{7},\Phi
_{6}\right\} }\,,\,\,d=-\frac{\left\{ \phi _{3}^{G},\Phi _{1}\right\} }{%
\left\{ \phi _{2}^{G},\Phi _{1}\right\} },\;w=-\frac{\left\{ \phi
_{4}^{G},\Phi _{1}\right\} }{\left\{ \phi _{2}^{G},\Phi _{1}\right\} }%
\,,\;z=-\frac{\left\{ \phi _{4}^{G},\Phi _{6}\right\} }{\left\{ \Phi
_{7},\Phi _{6}\right\} }\,,\;s=-\frac{\left\{ \phi _{4}^{G},\Phi
_{7}\right\} }{\left\{ \Phi _{6},\Phi _{7}\right\} }.
\end{eqnarray*}
The matrix $\left\{ \Phi _{a},\Phi _{b}\right\} $ has now a simple
quasi-diagonal form with the following nonzero elements\ 
\begin{eqnarray}
&&\left\{ \Phi _{2},\Phi _{1}\right\} =-\left\{ \Phi _{1},\Phi _{2}\right\}
=1\,,\;\left\{ \Phi _{3},\Phi _{3}\right\} =\left\{ \Phi _{4},\Phi
_{4}\right\} =\left\{ \Phi _{5},\Phi _{5}\right\} =-2i\,,\;  \nonumber \\
&&\left\{ \Phi _{6},\Phi _{7}\right\} =\left\{ \Phi _{7},\Phi _{6}\right\}
=2i\left[ \zeta (\tilde{\omega}_{0}+m)+\xi _{\bar{b}}\omega _{j}^{\bar{b}%
0}\left( \pi ^{j}-i\xi ^{j}\right) \right] \,,\;\left\{ \Phi _{8},\Phi
_{9}\right\} =\left\{ \Phi _{9},\Phi _{8}\right\} =1,  \nonumber \\
&&\left\{ \Phi _{10},\Phi _{11}\right\} =-\left\{ \Phi _{11},\Phi
_{10}\right\} =1\,,\;\left\{ \Phi _{12},\Phi _{12}\right\} =-\left\{ \Phi
_{13},\Phi _{13}\right\} =2i\,\,.  \label{15}
\end{eqnarray}

We call the variables ${\mbox{\boldmath$\y$\unboldmath}}=\left(
x^{k},p_{k},\zeta ,\xi ^{\bar{a}},\pi _{\bar{a}}\right) $ the independent
variables, since the remaining variables can be expressed via the
independent ones by means of the constraints. Similar to the flat space-time
case, we can prove that the Hamilton equations of motion and the
corresponding constraints for the independent variables have the form 
\begin{equation}
\dot{\mbox{\boldmath$\y$\unboldmath}}=\{{\mbox{\boldmath$\y$\unboldmath}},%
{\cal H}_{eff}\}_{D(U)}\,,\;U=\phi _{3,k}^{(1)}=0\,\,,\;\ k=1,2,3\,.
\label{16}
\end{equation}
The effective Hamiltonian ${\cal H}_{eff}$ reads: 
\begin{eqnarray}
&&{\cal H}_{eff}=\left[ \zeta \overline{{\cal A}}_{0}+\omega \right]
_{x^{0}=\zeta \tau }\,,\;\overline{{\cal A}}_{\mu }=qA_{\mu }+i\xi _{\bar{a}%
}\xi _{\bar{b}}\omega _{\mu }^{\bar{b}\bar{a}}\,,\ \omega =\left. \tilde{%
\omega}\right| _{\pi _{\bar{a}}=-i\xi _{\bar{a}}}=\sqrt{g_{00}\left( \omega
_{0}^{2}+\rho \right) }\;,  \nonumber \\
&&\omega _{0}=\sqrt{\left[ m^{2}-\left( p_{k}+\overline{{\cal A}}_{k}\right)
g^{kl}\left( p_{l}+\overline{{\cal A}}_{l}\right) -2iqF_{kl}\xi ^{k}\xi ^{l}%
\right] }\;,\;\;  \nonumber \\
\; &&\rho =4\left[ qF_{k\mu }e_{0}^{\mu }\xi ^{k}\bar{\pi}^{0}+\bar{\pi}%
_{0}\xi _{\bar{b}}\omega _{k}^{\bar{b}0}g^{kl}\left( p_{l}+\overline{{\cal A}%
}_{l}\right) \right] ,\;\bar{\pi}_{0}=-i\xi ^{k}\left( p_{k}+\overline{{\cal %
A}}_{k}\right) \left[ \zeta (\omega _{0}+m)\right] ^{-1},  \label{17}
\end{eqnarray}
and the only nonzero Dirac brackets between the independent variables are 
\begin{equation}
\left\{ x^{k},p_{l}\right\} _{D(U)}=\left\{ x^{k},p_{l}\right\} =\delta
_{l\,,}^{k}\;\;\left\{ \xi ^{\bar{a}},\xi ^{\bar{b}}\right\} _{D(U)}=\frac{i%
}{2}\eta ^{\bar{a}\bar{b}}\;.  \label{18}
\end{equation}

\section{Quantization}

Equal time commutation relations for the operators $\hat{X}^{k},\,\hat{P}%
_{k},\,\hat{\zeta},\,\hat{\Xi}^{\bar{a}},$ which correspond to the variables 
$x^{k},p_{k},\zeta ,$,$\xi ^{\bar{a}},$ are defined according to their Dirac
brackets. Thus, the nonzero commutators (anticommutators) are\thinspace\ 
\begin{equation}
\ [\hat{X}^{k},\hat{P}_{j}]=i\hbar \delta _{j}^{k}\,,\;\;[\hat{\Xi}^{\bar{a}%
},\hat{\Xi}^{\bar{b}}]_{+}=-\frac{\hbar }{2}\eta ^{\bar{a}\bar{b}%
}\,,\;k,j=1,2,3;\;\bar{a},\bar{b}=1,2,3\,.  \label{19}
\end{equation}
We assume $\hat{\zeta}^{2}=1$ (see \cite{GavGi00}), and realize the operator
algebra in the state space ${\cal R}$ whose elements $\mbox{\boldmath$\Psi$%
\unboldmath}\in {\cal R}$ are ${\bf x}$-dependent eight-component columns $%
\mbox{\boldmath$\Psi$\unboldmath}=\left( \Psi _{+1}({\bf x}),\Psi _{-1}({\bf %
x})\right) ,$ where $\Psi _{\zeta }({\bf x}),\;\zeta =\pm 1$ are four
component columns. The inner product in ${\cal R}$ is defined as follows: 
\begin{equation}
\left( \mbox{\boldmath$\Psi$\unboldmath},\mbox{\boldmath$\Psi$\unboldmath}%
^{\prime }\right) =\left( \Psi _{+1},\Psi _{+1}^{\prime }\right) _{D}+\left(
\Psi _{-1}^{\prime },\Psi _{-1}\right) _{D}\,\,.  \label{20}
\end{equation}
For the inner product between the four component columns we select the
following equivalent expressions, 
\begin{eqnarray}
&&\left( \Psi ,\Psi ^{\prime }\right) _{D}=\int \Psi ^{\dagger }({\bf x}%
)\Psi ({\bf x})g_{00}^{-1/2}\sqrt{-g}d{\bf x}\;=\int \overline{\Psi }({\bf x}%
)e_{a}^{0}(x)\gamma ^{a}\Psi ({\bf x})\sqrt{-g}d{\bf x}  \nonumber \\
&&{\bf =}\int \overline{\Psi }(x)\gamma ^{\mu }(x)\Psi (x)d\sigma _{\mu
}\;,\;\overline{\Psi }({\bf x})=\Psi ^{\dagger }({\bf x})\gamma
^{0},\;\gamma ^{\mu }(x)=e_{a}^{\mu }(x)\gamma ^{a},\;g=\det \left| \left|
g_{\mu \nu }\right| \right| \,.  \label{21}
\end{eqnarray}
To obey the above operator algebra in the space ${\cal R},$ we can chose the
following realization\footnote{%
Here and in what follows we use the following notations 
\[
{\rm bdiag}\left( A,\;B\right) =\left( 
\begin{array}{cc}
A & 0 \\ 
0 & B
\end{array}
\right) \,, 
\]
where $A$ and $B$ are some matrices.}: 
\begin{eqnarray}
&&\hat{X}^{k}=x^{k}{\bf I}\;,\;\;\;\hat{P}_{k}=\hat{p}_{k}{\bf I}\;,\;\;\hat{%
p}_{k}=-i\hbar \partial _{k}\,,  \nonumber \\
\; &&\hat{\Xi}^{\bar{a}}={\rm bdiag\,}(\hat{\xi}^{\bar{a}},\;\hat{\xi}^{\bar{%
a}})\,,\;\hat{\xi}^{\bar{a}}=\frac{i}{2}\hbar ^{1/2}\gamma ^{\bar{a}}\,\,,\;%
\hat{\zeta}={\rm bdiag\,}\left( I,\;-I\right) ,  \label{22}
\end{eqnarray}
where ${\bf I}$ is $8\times 8$ unit matrix, $I$ is $4\times 4$ unit matrix,
\ and $\gamma ^{\bar{a}},$ $\bar{a}=1,2,3,$ are three usual $\gamma $
-matrices in ($3+1)-$dimensions, $\;\left[ \gamma ^{\bar{a}},\gamma ^{\bar{b}%
}\right] _{+}=2\eta ^{\bar{a}\bar{b}}\,$ . One can easily see that such
defined operators are Hermitian with respect to the inner product (\ref{21}).

The quantum Hamiltonian $\hat{H}_{\tau }$ that defines the $\tau $-evolution
of state vectors of the system has to be constructed as a quantum operator
in the space ${\cal R}$ on the base of the correspondence principle starting
with its classical image, which is ${\cal H}_{eff}$ given by Eq. (\ref{17}).
There exist many quantum operators, which have the same classical image.
That corresponds to the well-known ambiguity of the general quantization. We
construct $\hat{H}_{\tau }$ as follows: 
\begin{eqnarray}
&&\hat{H}_{\tau }=\hat{\zeta}\widehat{{\cal A}}_{0}+\hat{\Omega}\,,\;%
\widehat{{\cal A}}_{0}=q\hat{A}_{0}+i\hat{\Xi}^{\bar{a}}\hat{\Xi}^{\bar{b}}%
\hat{\omega}_{0}^{\bar{b}\bar{a}},\;\hat{A}_{0}={\rm bdiag}\left( \left.
A_{0}\right| _{x^{0}=\tau }\,I,\;\;\left. A_{0}\right| _{x^{0}=-\tau
}\,I\right) ,  \nonumber \\
&&\hat{\omega}_{0}^{\bar{b}\bar{a}}={\rm bdiag}\left( \left. \omega _{0}^{%
\bar{b}\bar{a}}\right| _{x^{0}=\tau }\,I,\;\;\left. \omega _{0}^{\bar{b}\bar{%
a}}\right| _{x^{0}=-\tau }\,I\right) ,\;\hat{\Omega}={\rm bdiag}\left(
\left. \hat{\omega}\right| _{x^{0}=\tau \,,\;}-\left. \hat{\omega}\right|
_{x^{0}=-\tau }\right) ,  \nonumber \\
&&\hat{\omega}=e_{0}^{a}\gamma _{a}\left[ m+\gamma ^{k}(x)\left( \hat{p}%
_{k}+qA_{k}-i\frac{\hbar }{4}\gamma _{\bar{a}}\gamma _{\bar{b}}\omega _{k}^{%
\bar{b}\bar{a}}\right) \right] \,.  \label{25}
\end{eqnarray}
where we have used the Dirac matrix $\gamma ^{0},\left( \gamma ^{0}\right)
^{2}=1,\;\;\left[ \gamma ^{0},\gamma ^{\bar{a}}\right] _{+}=0.$ (One ought
to remark that we can write $e_{0}^{a}\gamma _{a}=\sqrt{g_{00}}\gamma _{0}$%
). The first term in the expression (\ref{25}) is a natural quantum image of
the classical quantity $\left. \zeta \overline{{\cal A}}_{0}\right|
_{x^{0}=\zeta \tau }$ . The term $\hat{\Omega}$ is a possible quantum image
of the classical quantity $\left. \omega \right| _{_{x^{0}=\zeta \tau }}$ .
In fact, we have to justify the following symbolic relation $\lim_{classical}%
\hat{\Omega}=\left. \omega \right| _{_{x^{0}=\zeta \tau }}\;.$ To be more
rigorous, one has to work with operator symbols. However, we remain here in
terms of the operators, hoping that our manipulations have a clear sense and
do not need to be confirmed on the symbol language. First, we replace the
operator $\hat{\Omega}$ under the sign of the limit by another one $\hat{%
\Omega}^{\prime }=\hat{\Omega}+\hat{\Delta}\,,\;$ 
\begin{eqnarray*}
\; &&\hat{\Delta}={\rm bdiag}\left( \left. \hat{\delta}\right| \Sb %
x^{0}=\tau  \\ \zeta =1  \endSb ,-\left. \hat{\delta}\right| \Sb x^{0}=-\tau 
\\ \zeta =-1  \endSb \right) \,,\;\hat{\delta}=\sqrt{g_{00}}\gamma _{0}\hat{%
\xi}^{k}\hat{\lambda}_{k}+i\frac{\hbar }{4}g^{jk}\partial _{0}g_{jk\,},\,%
\hat{\lambda}_{k}=-m\hbar ^{-1/2}\left[ \hat{y},\hat{\xi}_{k}\right] , \\
\;\; &&\hat{y}=\hat{\zeta}\left[ qF_{k\mu }e_{0}^{\mu }\hat{\xi}^{k}-\frac{1%
}{2}\left[ \hat{\xi}_{\bar{b}}\omega _{k}^{\bar{b}0}g^{kl},\hat{p}_{l}+%
\widehat{{\cal A}}_{l}\right] _{+},\frac{1}{\left( \hat{\omega}%
_{0}^{2}-m^{2}\right) }\right] _{+},\;\widehat{{\cal A}}_{k}=qA_{k}+i\hat{\xi%
}_{\bar{a}}\hat{\xi}_{\bar{b}}\omega _{k}^{\bar{b}\bar{a}}\,,
\end{eqnarray*}
since the classical limit of $\hat{\Delta}$ is zero. Indeed, the leading
contributions in $\hbar $ to the operator $\hat{\Delta}$ result from terms
that contain $\left( \hat{\xi}^{k}\right) ^{2}$. In the classical limit such
terms turn out to be proportional to $\left( \xi ^{k}\right) ^{2}$, which is
zero due to the Grassmann nature of $\xi $'s . The square of the operator $%
\hat{\Omega}^{\prime }$ in the classical limit corresponds to the square of
the classical quantity $\left. \omega \right| _{_{x^{0}=\zeta \tau }}\;.$
Indeed, $\left( \hat{\Omega}^{\prime }\right) ^{2}={\rm bdiag}\left( \left. 
\hat{\omega}^{2}\right| _{x^{0}=\tau ,\zeta =1},\left. \hat{\omega}%
^{2}\right| _{x^{0}=-\tau ,\zeta =-1}\right) ,\;\hat{\omega}%
^{2}=g_{00}\left( \hat{\omega}_{0}^{2}+\hat{\rho}_{1}+\hat{\rho}_{2}\right)
\,,\;$where 
\begin{eqnarray*}
&&\hat{\omega}_{0}^{2}=\left[ m^{2}-\frac{1}{\sqrt{-\det g_{ij}}}\left( \hat{%
p}_{k}+\widehat{{\cal A}}_{k}\right) \sqrt{-\det g_{ij}}g^{kl}\left( \hat{p}%
_{l}+\widehat{{\cal A}}_{l}\right) -\frac{\hbar ^{2}}{4}\bar{R}\right]
I-iqF_{jl}[\hat{\xi}^{j},\hat{\xi}^{l}]\;,\; \\
&\ &\hat{\rho}_{1}=\frac{1}{2i}\left[ \hat{y},\left[ \hat{\xi}^{l}\left( 
\hat{p}_{l}+\widehat{{\cal A}}_{l}\right) ,\hat{\omega}_{0}-m\right] _{+}%
\right] \;, \\
\; &&\hat{\rho}_{2}=\frac{m}{2i}\left\{ \left( \left[ \hat{\xi}^{l},\hat{y}%
\right] _{+}-\left[ g^{kl},\hat{y}\right] \hat{\xi}_{k}\right) \left( \hat{p}%
_{l}+\widehat{{\cal A}}_{l}\right) +g^{kl}\left[ \hat{p}_{l}+\widehat{{\cal A%
}}_{l},\hat{\xi}_{k}\right] \hat{y}-g^{kl}\left[ \left( \hat{p}_{l}+\widehat{%
{\cal A}}_{l}\right) \hat{y},\hat{\xi}_{k}\right] _{+}\right\} \\
&&+2i\hbar ^{-1/2}\left\{ \frac{1}{2}\left[ \hat{\xi}^{k},\hat{\xi}^{l}%
\right] \left[ \hat{\lambda}_{k},\hat{p}_{l}+\widehat{{\cal A}}_{l}\right] +%
\hat{\xi}^{k}\left[ \hat{\lambda}_{k},\hat{\xi}^{l}\right] \left( \hat{p}%
_{l}+\widehat{{\cal A}}_{l}\right) +\hat{\xi}^{l}\left[ \hat{p}_{l}+\widehat{%
{\cal A}}_{l},\hat{\xi}^{k}\right] \hat{\lambda}_{k}\right\} -\left( \hat{\xi%
}^{k}\hat{\lambda}_{k}\right) ^{2},
\end{eqnarray*}
where $\bar{R}$ is scalar curvature related to the stationary metric $\bar{g}%
_{\mu \nu }=\left. g_{\mu \nu }\right| _{x^{0}={\rm const}}$, and the
following expression for $\hat{\omega}$ is used, 
\begin{eqnarray*}
&&\hat{\omega}=\sqrt{g_{00}}\left( \hat{\omega}_{0}+i\frac{\hbar }{2}\gamma
^{k}(x)\gamma _{\bar{b}}\omega _{k}^{\bar{b}0}\right) =\sqrt{g_{00}}\hat{%
\omega}_{0}-i\frac{\hbar }{4}g^{jk}\partial _{0}g_{jk}\,,\,\,\bar{D}%
_{k}=\partial _{k}+\frac{1}{4}\gamma _{\bar{a}}\gamma _{\bar{b}}\omega _{k}^{%
\bar{b}\bar{a}}, \\
\; &&\hat{\omega}_{0}=\gamma _{0}\left[ m+\gamma ^{k}(x)\left( -i\hbar \bar{D%
}_{k}+qA_{k}\right) \right] =\gamma _{0}\left[ m-2i\hbar ^{-1/2}\hat{\xi}%
^{k}\left( \hat{p}_{k}+\widehat{{\cal A}}_{k}\right) \right] .
\end{eqnarray*}
In the classical limit $\hat{\omega}_{0}^{2}\rightarrow \omega _{0}^{2},$ $%
\hat{\rho}_{1}\rightarrow \rho ,$ $\hat{\rho}_{2}\rightarrow 0,$ ($\hat{\rho}%
_{2}$ does not contain terms without $\hbar $). Thus, the classical limit of 
$\left( \hat{\Omega}^{\prime }\right) ^{2},$ and therefore $\left( \hat{%
\Omega}\right) ^{2}$ as well, is the classical quantity $\left. \omega
^{2}\right| _{x^{0}=\zeta \tau \,}$.

The Hamiltonian (\ref{25}) can be written in the following block-diagonal
form 
\begin{equation}
\hat{H}_{\tau }={\rm bdiag}\left( \hat{h}(\tau ),-\hat{h}(-\tau )\right) ,\;%
\hat{h}(x^{0})=qA_{0}-i\frac{\hbar }{4}\gamma _{\bar{a}}\gamma _{\bar{b}%
}\omega _{0}^{\bar{b}\bar{a}}+\hat{\omega}\;.\;  \label{27}
\end{equation}
The $\tau $-evolution of the state vectors is defined by the corresponding
Schr\"{o}dinger equation $i\hbar \partial _{\tau }\mbox{\boldmath$\Psi$%
\unboldmath}(\tau )=\hat{H}_{\tau }\mbox{\boldmath$\Psi$\unboldmath}(\tau ),$
where the state vectors now depend parametrically on $\tau $, $%
\mbox{\boldmath$\Psi$\unboldmath}(\tau )=\left( \Psi _{+1}(\tau ,{\bf x}%
),\Psi _{-1}(\tau ,{\bf x})\right) $. Similar to the flat space-time case 
\cite{GavGi00}, we may reformulate the evolution in terms of the physical
time $x^{0}=\zeta \tau .$ The corresponding Schr\"{o}dinger equation has the
form 
\begin{eqnarray}
&&i\hbar \partial _{0}\mbox{\boldmath$\Psi$\unboldmath}(x^{0})=\hat{H}%
_{x^{0}}\mbox{\boldmath$\Psi$\unboldmath}(x^{0}),\;\hat{H}_{x^{0}}={\rm bdiag%
}\left( \hat{h}(x^{0})\,,\;\ \hat{h}^{c}(x^{0})\right) \;,  \nonumber \\
&&\hat{h}^{c}(x^{0})=\gamma ^{2}\left( \hat{h}(x^{0})\right) ^{\ast }\gamma
^{2}=\left. \hat{h}(x^{0})\right| _{q\rightarrow -q},\;\mbox{\boldmath$\Psi$%
\unboldmath}(x^{0})=\left( \Psi (x),\Psi ^{c}(x)\right) ,  \label{28}
\end{eqnarray}
where $\Psi (x)=$ $\Psi _{+1}(x^{0},{\bf x})$ and $\Psi ^{c}(x)=\gamma
^{2}\Psi _{-1}^{\ast }(-x^{0},{\bf x})$ are Dirac bispinors. The inner
product of two states vectors in such a representation reads 
\begin{equation}
\left( \mbox{\boldmath$\Psi$\unboldmath},\mbox{\boldmath$\Psi$\unboldmath}%
^{\prime }\right) =\left( \Psi ,\Psi ^{\prime }\right) _{D}+\left( \Psi ^{c},%
{\Psi ^{c}}^{\prime }\right) _{D}\;.  \label{29}
\end{equation}
In accordance with the classical interpretation (see \cite{GavGi00}) we
regard $\hat{\zeta}$ as the charge sign operator. Let $\mbox{\boldmath$\Psi$%
\unboldmath}_{\zeta }$ be states with a definite charge, thus, $\hat{\zeta}%
\mbox{\boldmath$\Psi$\unboldmath}_{\zeta }=\zeta \mbox{\boldmath$\Psi$%
\unboldmath}_{\zeta }\,,\;\;\zeta =\pm 1\;.$ The states $\mbox{\boldmath$%
\Psi$\unboldmath}_{+1}$ have $\Psi ^{c}=0$. Then (\ref{28}) is reduced to
the Dirac equation in curved space-time for the spinor field $\Psi (x)$ of
the charge $q$, 
\begin{equation}
\left[ \gamma ^{\mu }\left( i\hbar D_{\mu }-qA_{\mu }\right) -m\right] \Psi
(x)=0,\;D_{\mu }=\partial _{\mu }+\frac{1}{4}\gamma _{a}\gamma _{b}\omega
_{\mu }^{ba}.  \label{30}
\end{equation}
States $\mbox{\boldmath$\Psi$\unboldmath}_{-1}$ have $\Psi =0$. Then (\ref
{28}) is reduced to the Dirac equation in curved space-time for the spinor
field $\Psi ^{c}(x)$ of the charge -$q$, 
\begin{equation}
\left[ \gamma ^{\mu }\left( i\hbar D_{\mu }+qA_{\mu }\right) -m\right] \Psi
^{c}(x)=0.  \label{31}
\end{equation}

The Hamiltonian $\hat{h}(x^{0})$ can be considered as a one-particle Dirac
Hamiltonian in the case under consideration for the charge $q$.

Let us restrict ourselves to those backgrounds that do not create particles
from the vacuum. For such backgrounds the one-particle sector of the
corresponding QFT can be consistently defined, see the corresponding
discussion in \cite{GavGi00} and some remarks at the end. Consider for
simplicity, the eigenvalue problem $\hat{h}\Psi ({\bf x})=\epsilon \Psi (%
{\bf x})$ for the Dirac Hamiltonian in a time-independent external
background (in fact, $\hat{h}(x^{0})$ does not depend on $x^{0}$ in such a
case, thus, $\hat{h}(x^{0})=\hat{h}$)$\,.$ Presenting the spinor $\Psi $ in
the form 
\[
\Psi ({\bf x})=\left[ g_{00}^{-1/2}\gamma ^{0}\left( \epsilon -qA_{0}+i\frac{%
\hbar }{4}\gamma _{a}\gamma _{b}\omega _{0}^{ba}\right) +\gamma ^{k}\left(
i\hbar D_{k}-qA_{k}\right) +m\right] \varphi ({\bf x})\;. 
\]
we get for $\varphi ({\bf x})$ the corresponding squared Dirac equation, 
\begin{eqnarray}
&&\left[ g^{00}\left( \epsilon -qA_{0}+i\frac{\hbar }{4}\gamma _{a}\gamma
_{b}\omega _{0}^{ba}\right) ^{2}-{\cal D}\right] \varphi ({\bf x})=0\,,\; 
\nonumber \\
&&{\cal D}=m^{2}-\frac{\hbar ^{2}}{4}R-\frac{1}{\sqrt{-g}}\left( i\hbar
D_{k}-qA_{k}\right) \sqrt{-g}g^{kl}\left( i\hbar D_{l}-qA_{l}\right) +i\frac{%
\hbar }{4}qF_{\mu \nu }[\gamma ^{\mu },\gamma ^{\nu }]_{-}\;,  \label{32}
\end{eqnarray}
where $R$ is the scalar curvature. We can see that a pair $(\varphi
,\;\epsilon )$ is a solution of the equation (\ref{32}) if it obeys either
the equation $\epsilon =qA_{0}-i\frac{\hbar }{4}\gamma _{a}\gamma _{b}\omega
_{0}^{ba}+\sqrt{\varphi ^{-1}g_{00}{\cal D}\varphi }\;,$ or the equation $%
\epsilon =qA_{0}-i\frac{\hbar }{4}\gamma _{a}\gamma _{b}\omega _{0}^{ba}-%
\sqrt{\varphi ^{-1}g_{00}{\cal D}\varphi }\;.$ \ Let us denote via $(\varphi
_{+,n},\;\epsilon _{+,n})$ the solutions of the first equation, and via $%
(\varphi _{-,n},\;\epsilon _{-,n})$ the solutions of the second equation,
where $n$ are quantum numbers. Thus, the eigenvalue problem has the
solutions 
\[
\epsilon _{+,n}=qA_{0}-i\frac{\hbar }{4}\gamma _{a}\gamma _{b}\omega
_{0}^{ba}+\sqrt{\varphi _{+,n}^{-1}g_{00}{\cal D}\varphi _{+,n}}%
,\,\,\,\epsilon _{-,\alpha }=qA_{0}-i\frac{\hbar }{4}\gamma _{a}\gamma
_{b}\omega _{0}^{ba}-\sqrt{\varphi _{-,\alpha }^{-1}g_{00}{\cal D}\varphi
_{-,\alpha }}, 
\]
and 
\begin{eqnarray}
\Psi _{+,n}({\bf x}) &=&\left[ g_{00}^{-1/2}\gamma ^{0}\left( \epsilon
_{+,n}-qA_{0}+i\frac{\hbar }{4}\gamma _{a}\gamma _{b}\omega _{0}^{ba}\right)
+\gamma ^{k}\left( i\hbar D_{k}-qA_{k}\right) +m\right] \varphi _{+,n}({\bf x%
})\;,  \nonumber \\
\Psi _{-,n}({\bf x}) &=&\left[ g_{00}^{-1/2}\gamma ^{0}\left( \epsilon
_{-,n}-qA_{0}+i\frac{\hbar }{4}\gamma _{a}\gamma _{b}\omega _{0}^{ba}\right)
+\gamma ^{k}\left( i\hbar D_{k}-qA_{k}\right) +m\right] \varphi _{-,n}({\bf x%
})\;,  \label{34}
\end{eqnarray}
Since $\epsilon _{+,n}>$ $\epsilon _{-,\alpha }\,,$ we call $\epsilon _{+,n}$
the upper branch and $\epsilon _{-,\alpha }$ the lower branch of the energy
spectrum. The square norm of the eigenvectors $\Psi _{\varkappa ,n}$ is
always positive, all the eigenvectors are mutually orthogonal and can thus
be orthonormalized as follows, 
\begin{equation}
\left( \Psi _{\varkappa ,n},\Psi _{\varkappa ^{\prime },n^{\prime }}\right)
_{D}=\delta _{\varkappa ,\varkappa ^{\prime }}\delta _{n,n^{\prime
}},\;\;\varkappa =\pm \;.  \label{35}
\end{equation}
A solution of the eigenvalue problem $\hat{h}^{c}\Psi ^{c}=\epsilon \Psi
^{c} $ for the charge conjugated Hamiltonian can be analyzed in a similar
manner. Here we get the set $\left( \epsilon _{\varkappa ,n}^{c},\Psi
_{\varkappa ,n}^{c}\right) ,\;$ 
\begin{equation}
\Psi _{\varkappa ,n}^{c}=\gamma ^{2}\Psi _{-\varkappa ,n}^{\ast
}\;,\;\;\epsilon _{\varkappa ,n}^{c}=-\epsilon _{-\varkappa ,n}\,,\;\left(
\Psi _{\varkappa ,n}^{c},\Psi _{\varkappa ^{\prime },n^{\prime }}^{c}\right)
=\delta _{\varkappa ,\varkappa ^{\prime }}\delta _{n,n^{\prime
}},\;\;\varkappa =\pm \;.  \label{36}
\end{equation}

>From this point on, we can repeat all the arguments from \cite{GavGi00} and
construct a consistent relativistic quantum mechanics in Hilbert space
without an indefinite metric, which is a reduction of the space ${\cal R}$
to its physical subspace. The latter can be defined as a linear envelope of
vectors of the form 
\[
\mbox{\boldmath$\Psi$\unboldmath}_{+,n}=\left( 
\begin{array}{c}
\psi _{+,n}({\bf x}) \\ 
0
\end{array}
\right) ,\;\mbox{\boldmath$\Psi$\unboldmath}_{+,\alpha }^{c}=\left( 
\begin{array}{c}
0 \\ 
\psi _{+,\alpha }^{c}({\bf x})
\end{array}
\right) \,. 
\]
In such a Hilbert space the operator $\hat{\Omega}$ has a positively defined
spectrum and the Hamiltonian $\hat{H}_{x^{0}}$ has the right spectrum of
particle and antiparticle energies in the background under consideration,
which coincides with the spectrum of particles and antiparticles in the
one-particle sector of the corresponding QFT.

\section{Some remarks}

Returning to our choice of the quantum Hamiltonian (in fact, of the operator 
$\hat{\Omega}\;$from (\ref{25})$),$ one has to stress that the classical
theory gives enough information to resolve the ordering problem in an unique
way. The operator ordering and the nonclassical parts of the operator $\hat{%
\Omega}$ were chosen to maintain the invariance of the quantum theory under
general coordinate transformations and under $U\left( 1\right) $
transformations of the electromagnetic background. In particular, such a
choice provides the invariance of the inner product (\ref{21}) under general
coordinate transformations as well as under the choice of the space-like
hypersurface where the inner product is defined. One can also see that $\hat{%
\Omega}$ is positive defined in the Hilbert space constructed. The
positivity condition helps to fix an ambiguity in the definition of $\hat{%
\Omega}$ as well.

We recall that a one-particle sector of QFT (as well as any sector with a
definite particle number) may be defined in an unique way for all time
instants only in external backgrounds which do not create particles from the
vacuum . Nonsingular time independent external backgrounds are important
examples of the above backgrounds. That is why we have presented the
detailed discussion for such kinds of backgrounds to simplify our analyses.
A generalization to arbitrary backgrounds, in which the vacuum remains
stable, may be done in a similar manner. In backgrounds that violate the
vacuum stability, a more complicated multi-particle interpretation of the
constructed quantum mechanics, which establishes a connection to the QFT, is
also possible.

{\bf Acknowledgment} The authors are thankful to the foundations FAPESP,
FAPESE (S.P.G.) and CNPq (D.M.G) for support. Besides, S.P.G. thanks the
Department of Nuclear Physics of USP for hospitality.

\end{document}